\documentclass[fleqn,10pt]{wlscirep}
\usepackage[utf8]{inputenc}
\usepackage[T1]{fontenc}
\title{Needle bevel geometry influences the flexural deflection magnitude in ultrasound-enhanced fine-needle biopsy}

\author[1]{Saif Bunni}
\author[1,*]{Heikki J. Nieminen}
\affil[1]{Medical Ultrasonics Laboratory (MEDUSA),
Department of Neuroscience and Biomedical Engineering (NBE), Aalto University School of Science, 02150, Espoo, Finland.}

\affil[*]{heikki.j.nieminen@aalto.fi}

\begin{abstract}

It has been recently demonstrated that use of ultrasound increases the tissue yield in ultrasound-enhanced fine-needle aspiration biopsy (USeFNAB) as compared to conventional fine-needle aspiration biopsy (FNAB). To date, the association between bevel geometry and needle tip action has not been widely explored. In this study, we studied the needle resonance characteristics and deflection magnitude of various needle bevel geometries with varying bevel lengths. With a conventional lancet, having a 3.9 mm long bevel, the tip deflection efficiency in air and water was 220 and 105 \textmu m/W, respectively. This was higher in comparison to an axi-symmetric tip, having a bevel length of 4 mm, which achieved a deflection efficiency of 180 and 80 \textmu m/W in air and water, respectively. This study emphasised the importance of relationship between flexural modulus of bevel geometry in the context of various insertion media and, thus, could provide understanding on approaches to control post-puncture cutting action by modifying the needle bevel geometry, essential for the USeFNAB application.

\end{abstract}
\begin{document}

\flushbottom
\maketitle
%
%
\thispagestyle{empty}

\section*{Introduction}
\label{sec:introduction}

Fine-needle aspiration biopsy (FNAB) is a method employing needles to obtain a tissue sample from a suspected pathology \cite{frable1983fine}\textsuperscript{,} \cite{zajicek1977aspiration}\textsuperscript{,} \cite{wu2004fine}. It has been shown that Franseen-type tips obtain higher diagnostic yield than the conventional lancet \cite{Asokkumar2019}, and a Menghini tip \cite{Bang2021}. Axi-symmetric (i.e. circumferential) bevels have also been suggested to increase the likelihood of a histo-pathologically adequate sample \cite{Greene1985}.

When operating the needle, the needle is penetrated through the skin and layers of tissue to access suspected pathology. Recent studies suggest that ultrasonic actuation could reduce the required puncture forces into soft tissue \cite{Sadiq2013}\textsuperscript{,} \cite{Yang2004}\textsuperscript{,} \cite{Cai2019}\textsuperscript{,} \cite{Tan2018}. Geometry of needle-bevel has been shown to influence needle-interaction forces, for example, longer bevel lengths have been shown to exhibit lower tissue-puncture forces \cite{Jiang2014}. After the needle has penetrated tissue surface i.e. post-puncture, it has been suggested that cutting forces of needle could contribute up to 75\% of the total needle-tissue interaction forces \cite{Okamura2004}. In post-puncture stages, it has been demonstrated that ultrasound (US) could increase diagnostic biopsy yield in soft tissues \cite{Perra2021}. On the other hand, methods with US-enhancement of biopsy of bone have been developed for sampling hard tissues \cite{li2017surgical}\textsuperscript{,} \cite{Mathieson2017}, but no biopsy yield results were reported. While there are many studies concerning the axial (longitudinal) static forces in needle-tissue interaction \cite{Abolhassani2007}\textsuperscript{,} \cite{van2012needle}, there has been limited research on temporal dynamics and needle bevel geometry in ultrasound-enhanced FNAB (USeFNAB).

The aim of this study was to investigate the role of different bevel geometries on needle tip action in a flexurally-actuated needle. More specifically, we studied in post-puncture, the influence of insertion-medium on needle tip deflection, for a conventional needle bevel (i.e. the lancet), axi-symmetric and a-symmetric single-step bevel geometries. Understanding how the needle-tip action is controlled could be beneficial in the development of ultrasound-enhanced biopsy needles for different purposes, such as selectively obtaining an aspirate or soft tissue cores.

\section*{Methods}

Our approach was to first model the change of flexural wavelength, with different bevel lengths in selected bevel geometries. We then simulated the effect of tube length on the transfer mechanical mobility for a conventional lancet, axi-symmetric and a-symmetric single-step bevel geometries, in order to identify the optimal combinations of tube and bevel lengths. Following this, the frequency was experimentally characterised by measuring the voltage reflection coefficients and calculating the power transfer efficiency of the needle constructs, in air, water and ballistic gelatin 10\% (w/v), from which the operational frequency was identified for prototyping. For the prototype needles, the flexural-wave deflection at the needle-tip was then measured in air and water, and the deflection efficiency and electrical power transmitted to insertion medium were estimated for each bevel geometry.

\subsection*{Flexural wavelength}

For an infinite (boundless) beam with a cross-sectional area $A$, the flexural (or bending) phase velocity $c_{EI}$ was defined \cite{Cremer1973}: 
\begin{equation}
    c_{EI} = \sqrt[4]{EI/m'}\sqrt{\omega_0},
\label{flexwave}
\end{equation}

\noindent where $E$ was the Young’s Modulus (N/m\textsuperscript{2}), $\omega_0 = 2\pi f_0$ was the excitation angular frequency (rad/s), where $f_0$ was the linear frequency (1/s or Hz), $I$ was the area moment of inertia (m\textsuperscript{4}) around the axis of interest, and $m'=\rho_0 A$ was the mass per unit length (kg/m), where $\rho_0$ was the density (kg/m\textsuperscript{3}), and $A$ was the cross-sectional ($xy$-plane) area of the beam (m\textsuperscript{2}). Since the force applied in our case, was parallel to the vertical $y$-axis i.e. $\Tilde{F}_y\Vec{j}$, we were only concerned with the area moment of inertia around the horizontal $x$-axis i.e. $I_{xx}$, hence:

\begin{equation}
    I_{xx} = \oint_A (y-y_{CG})^2 \,dA,
\end{equation}

\noindent where $y_{CG}$ is the $y$-coordinate of the centre of gravity of the needle tube in the $xy$-plane, as illustrated in Fig.~\ref{fig:modelsetup}.

\subsection*{Finite element model (FEM)}

Assuming a purely harmonic displacement, acceleration was expressed as $\partial^2 \Vec{u}/\partial t^2 = -\omega^2\Vec{u}$, such as that $\Vec{u}(x, y, z, t) := u_x\Vec{i} + u_y\Vec{j}+ u_z\Vec{k}$ was a three-dimensional displacement vector (dimensionless) defined in the spatial coordinates. Substituting the latter, the law of balance of momentum in its Lagrangian form for finite deformation \cite{Reddy2013}, was given according to its implementation in COMSOL Multiphysics software (version 5.4--5.5, COMSOL Inc., Massachusetts, USA), as:

\begin{equation}
    -\rho_0\omega_0^2\Vec{u}=\Vec{\nabla}\cdot\underline{\sigma} + \Vec{F_V}e^{j\phi},
\label{fem}    
\end{equation}

\noindent where $\Vec{\nabla}:= \partial/\partial x\Vec{i} + \partial/\partial y\Vec{j} + \partial/\partial z\Vec{k}$ was the tensor divergence operator, and $\underline{\sigma}$ was the second Piola-Kirchhoff stress tensor (of second order, N/m\textsuperscript{2}), and $\Vec{F_V}:= F_{V_x}\Vec{i}+ F_{V_y}\Vec{j}+ F_{V_z}\Vec{k}$ was the volumetric force vector per deformed volume (N/m\textsuperscript{3}), and $e^{j\phi}$ was the phase of the volumetric force having a phase angle $\phi$ (rad). In our case, the volumetric body force was zero, and our model assumed geometrical linearity, and small purely elastic strain i.e. $\underline{\varepsilon}^{el} = \underline{\varepsilon}$, where $\underline{\varepsilon}^{el}$ and $\underline{\varepsilon}$ were the elastic and total strains, respectively (of second order, dimensionless). The Constitutive \textit{Hookean} isotropic elasticity tensor $\underline{\underline{C}}$ was defined using the Young’s Modulus $E$ (N/m\textsuperscript{2}) and the Poisson’s ratio $v$, so that $\underline{\underline{C}}:=\underline{\underline{C}}(E,v)$ (of fourth order). Therefore the calculation of stress becomes $\underline{\sigma} := \underline{\underline{C}}:\underline{\varepsilon}$.

Computation was done with 10-node tetrahedral elements with element size of $\leq$ 8 \textmu m. The needle was simulated in a vacuum, and a magnitude of transfer mechanical mobility (m$\cdot$s$^\textnormal{-1}\cdot$N$^\textnormal{-1}$) was defined as $\lvert\Tilde{Y}_{v_yF_y}\rvert = \lvert\Tilde{v}_y\Vec{j}\rvert/\lvert\Tilde{F}_y\Vec{j}\rvert$ \cite{Fahy2007}, where $\Tilde{v}_y\Vec{j}$ was the the output complex velocity at the tip, and $\Tilde{F}_y\Vec{j}$ was the complex driving force located at the proximal end of the tube, as illustrated in Fig.~\ref{fig:modelsetup}(b).

\subsection*{Fabrication of needle constructs}
\label{sec:needleconstructs}

\begin{figure}
    \centering
    \includegraphics[scale=1]{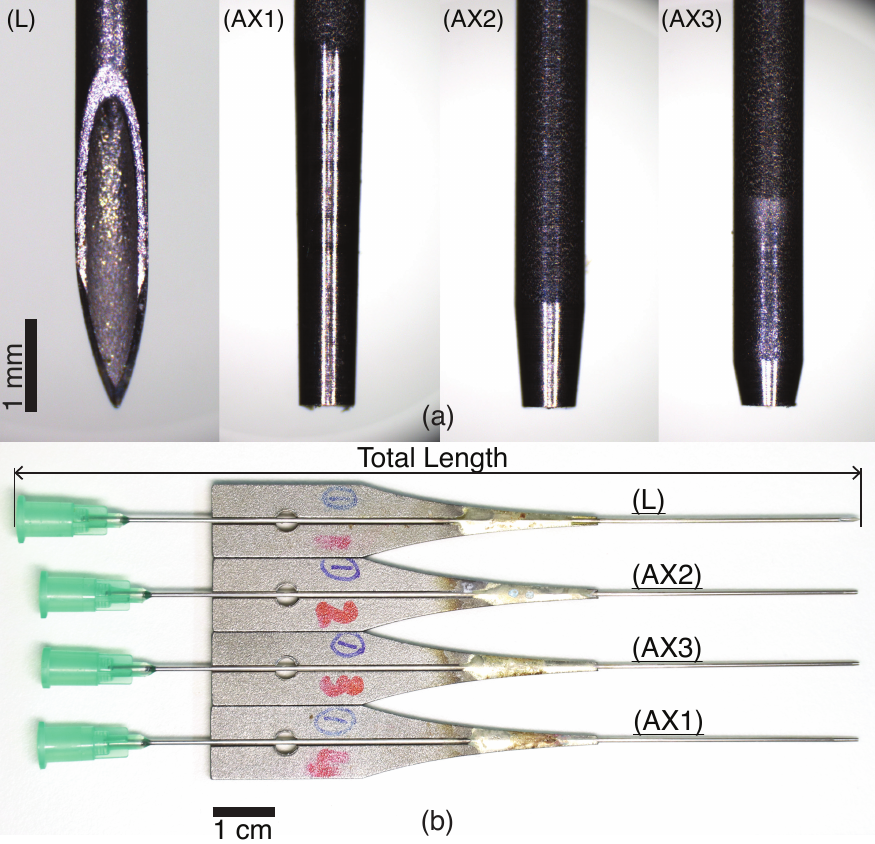}
    \caption{\textbf{(a)} Characterisation was done for four needle tip types (photographed), a commercially-available lancet (L), and three fabricated axi-symmetric single-step bevels (AX1--3), with bevel lengths (BL) of 4, 1.2 and 0.5 mm, respectively. \textbf{(b)} Top view of the four needle constructs (photographed), which were were then attached to the transducer via a M4 bolt.}
    \label{fig:bevels}
\end{figure}

The needle constructs (Fig.~\ref{fig:bevels}) consisted of a conventional 21 Gauge hypodermic needle (catalogue number: 4665643, Sterican\textsuperscript{\textregistered}, outer diameter 0.8 mm, length 120 mm, stainless chromium nickel steel AISI type 304 grade, B. Braun Melsungen AG, Melsungen, Germany) fitted with a Luer Lock plastic hub made of polypropylene at the proximal end, and modified accordingly at the tip. Needle tubes were soldered to waveguides, as shown in Fig.~\ref{fig:bevels}(b). The waveguides were 3D printed with stainless steel (EOS Stainless Steel 316L in EOS M 290 3D Printer, 3D Formtech Oy, Jyväskylä, Finland), then fastened via a M4 bolt to a Langevin transducer. The Langevin transducer consisted of 8 piezo ring elements, loaded by two masses at either end.

\begin{figure}
    \centerline{\includegraphics[scale=1]{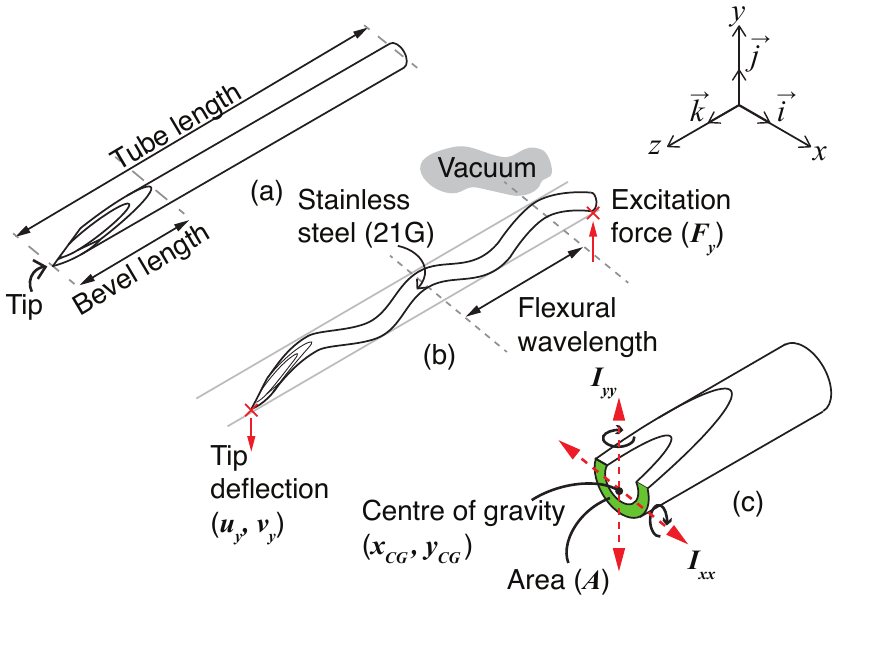}}
    \caption{Definition of the flexural wavelength, and setup of finite element model (FEM) of needle and boundary conditions. \textbf{(a)} Definition of the total tube length tube length (TL) and bevel length (BL). \textbf{(b)} A 3-dimensional (3D) finite element model (FEM) employed a harmonic point force $\Tilde{F}_y\Vec{j}$ to excite the needle tube at the proximal end, a point deflection and velocity ($\Tilde{u}_y\Vec{j}$, $\Tilde{v}_y\Vec{j}$) was measured at the tip to allow a calculation of transfer mechanical mobility. $\lambda_y$ was defined as the flexural wavelength associated with the vertical force $\Tilde{F}_y\Vec{j}$. \textbf{(c)} Definitions of the centre of gravity, the cross-sectional area $A$, and the moments of inertia around the $x$ and $y$ axes.}
\label{fig:modelsetup}
\end{figure}

Three axi-symmetrically bevelled tips were fabricated (Fig.~\ref{fig:bevels}) (TAs Machine Tools Oy) with bevel lengths (BL, as defined in Fig.~\ref{fig:modelsetup}(a)) of 4.0, 1.2 and 0.5 mm, corresponding to bevel angles (BA) of $\approx$ 2$^\circ$, 7$^\circ$, and 18$^\circ$, respectively. The masses of waveguides and needles were 3.4 $\pm$ 0.017 g (mean $\pm$ s.d., $n$ = 4) for bevels L and AX1--3, respectively (Quintix\textsuperscript{\textregistered} 224 Design 2, Sartorius AG, Göttingen, Germany). The total lengths from needle tip to the end of the plastic hub were 13.7, 13.3, 13.3, 13.3 cm, for bevels L and AX1--3 in Fig.~\ref{fig:bevels}(b), respectively.

For all needle constructs, the length from needle tip to the tip of the waveguide was 4.3 cm, and the needle tube was orientated so that bevel planes faced upwards (i.e. parallel to the $y$-axis), as in (Fig.~\ref{fig:modelsetup}).

\subsection*{Modal analysis}

A custom-script in MATLAB (R2019a, The MathWorks Inc., Massachusetts, USA), running on a computer (Latitude 7490, Dell Inc., Texas, USA), was used to generate a linear sine-sweep from 25 to 35 kHz for duration of 7 seconds, which was converted to an analogue signal via a digital-to-analogue (DA) converter (Analog Discovery 2, Digilent Inc., Washington, USA). The analogue signal $V_0$ (0.5 V\textsubscript{pk-pk}) was then amplified using a custom-made radio frequency (RF) amplifier (Mariachi Oy, Turku, Finland). The incident amplified voltage ${V_I}$ was output from the RF amplifier at an output impedance of 50 $\Omega$, to the transformer built into the needle construct, which had an input impedance of 50 $\Omega$. The Langevin transducer (back and front mass-loaded sandwich piezoelectric transducer) was used to generate the mechanical wave. The custom-made RF amplifier was equipped with a dual-channel standing-wave power ratio (SWR) meter, which allowed both the incident ${V_I}$ and reflected amplified voltages $V_R$ to be recorded via the analogue-digital (AD) converters (Analog Discovery 2) at sampling frequency of 300 kHz. The excitation signal was amplitude modulated at the beginning and end to prevent signal transients overloading the amplifier’s input.

\begin{figure*}
    \centering
    \includegraphics[scale=1]{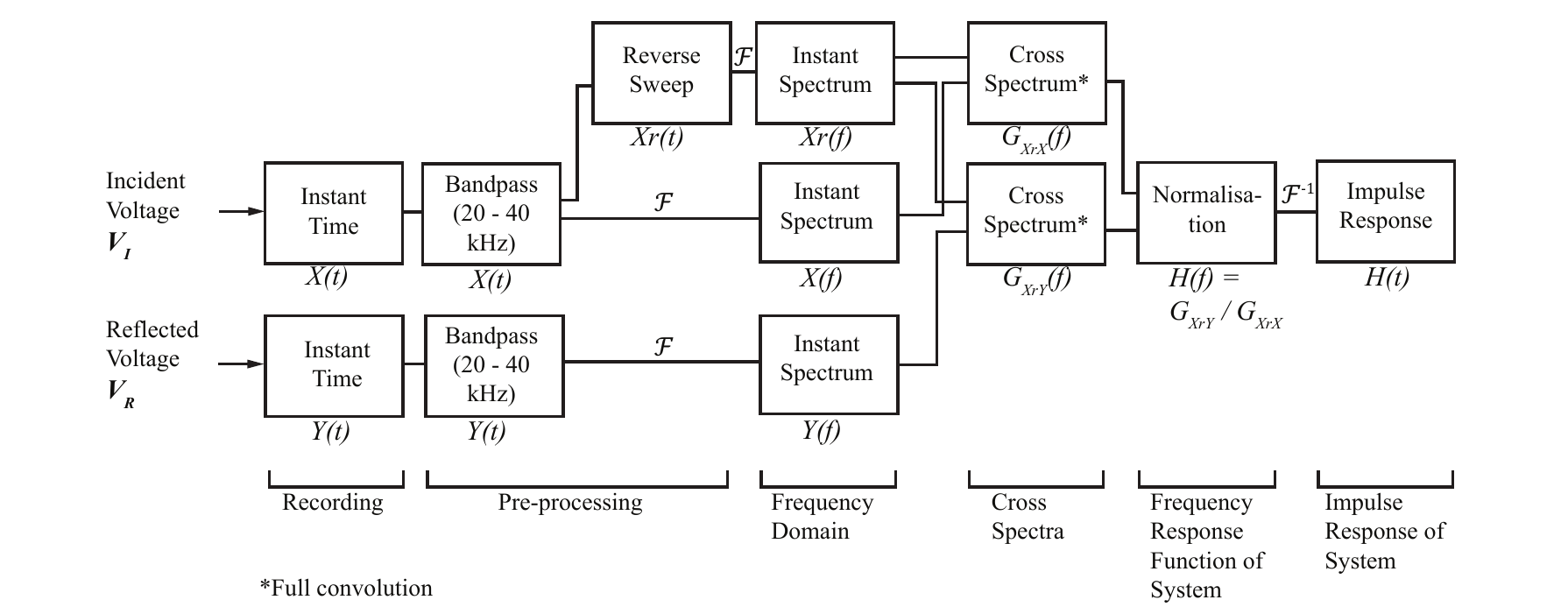}
    \caption{Measurement method of the frequency response functions (FRFs) of needle constructs. Swept-sine dual-channel measurement \cite{Poletti1988}\textsuperscript{,} \cite{Herlufsen1984} was used to obtain frequency response functions $\Tilde{H}(f)$ and its impulse responses $H(t)$. $\mathcal{F}$ and $\mathcal{F}^{-1}$ denote a digital truncated Fourier transform operation and its inverse, respectively. $G(f)$ denotes multiplication of two signals in the frequency domain, e.g. $G_{XrX}$ means a multiplication of the reverse sweep $Xr(f)$ and the incident voltage $X(f)$ signals, respectively.}
\label{fig:dspsetup}
\end{figure*}

Using a custom script implemented in MATLAB, frequency response functions (FRFs) i.e. $\Tilde{H}(f)$, were estimated offline using a swept-sine dual channel measurement technique \cite{Poletti1988} (Fig.~\ref{fig:dspsetup}), which assumed a linear time-invariant system. In addition, a bandpass filter having passband between 20 and 40 kHz was applied to remove any unwanted frequencies from signal. In reference to transmission line theory, $\Tilde{H}(f)$ in this case was equivalent to the voltage reflection coefficient i.e. $\rho_{V} \equiv {V_R}/{V_I}$ \cite{ramo1965fields}. Since the amplifier output impedance $Z_0$ was matched to input impedance of the transformer built-in with the transducer, the electrical power reflection coefficient ${P_R}/{P_I}$ was reduced to ${V_R}^2/{V_I}^2$ i.e. $\lvert\rho_{V}\rvert^2$. In the case when absolute values of electrical power were needed, the incident $P_I$ and reflected $P_R$ powers (W) were calculated by taking the root-mean-square (r.m.s.) of the corresponding voltages, such as that for a sinusoidally-excited transmission line, $P = {V}^2/(2Z_0)$ \cite{ramo1965fields}, where $Z_0$ was 50 $\Omega$. The electrical power transmitted to the load $P_T$ (i.e. to the insertion medium) could be calculated as $\lvert P_I - P_R \rvert$, and the power transfer efficiency (PTE) could be defined and given as a percentage (\%), so that \cite{otung2021communication}: 
\begin{equation}
    \textnormal{PTE} = \frac{\lvert P_I - P_R \rvert}{P_I} = (1 - \lvert\rho_{V}\rvert^2)*100.
\label{alpha_t}
\end{equation}

The FRFs were then used to estimate the modal frequencies $f_{1-3}$ (kHz) of the needle construct, and their corresponding power transfer efficiencies PTE\textsubscript{1--3}. The full-width at half-maxima (FWHM\textsubscript{1--3}, Hz) were estimated directly from PTE\textsubscript{1--3}, obtained from the one-sided linear frequency spectra at modal frequencies $f_{1-3}$ described in Table.~\ref{tab:ModalRegions}.

\subsection*{Needle deflection measurement}

\begin{figure}
    \centerline{\includegraphics[scale=1]{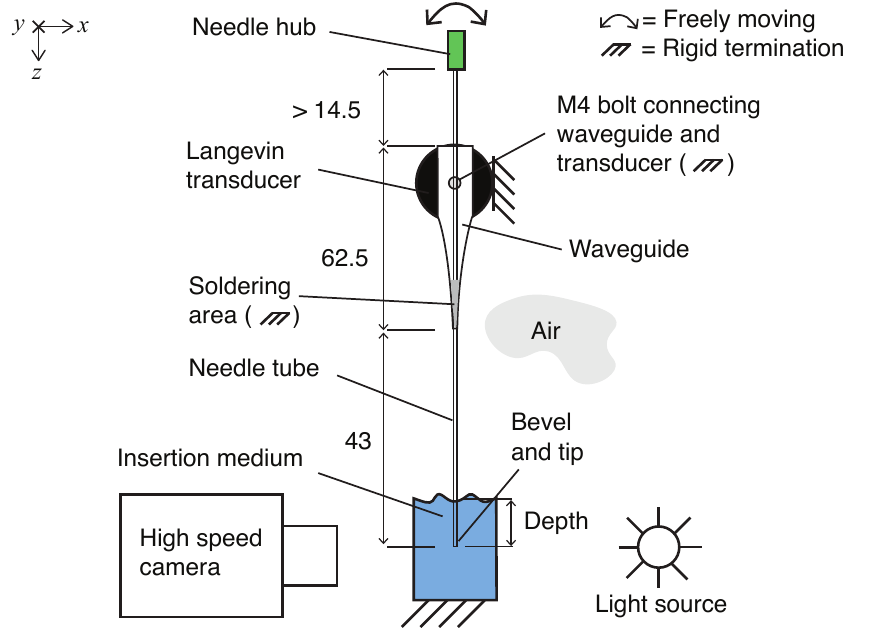}}
    \caption{Front-view of experimental setup. Depth was measured from surface of medium. The needle construct was clamped and mounted on a motorised translation stage. A high speed camera with a high magnification (5$\times$) lens was used to measure the deflection of bevel-tip. All dimensions are given in mm.}
\label{fig:expsetup_cam}
\end{figure}

As shown in Fig.~\ref{fig:expsetup_cam}, a high speed camera (Phantom V1612, Vision Research Inc., New Jersey, USA), fitted with a macro lens (MP-E 65 mm, $\mathit{f}$/2.8, 1--5$\times$, Canon Inc., Tokyo, Japan), was used to record the deflection of the needle tip undergoing flexural excitation (single frequency, continuous sinusoid) at frequencies 27.5--30 kHz. In order to produce shadowgraphs, a cooled high-intensity white LED element was placed behind the needle bevel (catalogue number: 4052899910881, White Led, 3000 K, 4150 lm, Osram Opto Semiconductors GmbH, Regensburg, Germany).

\begin{figure*}
    \centering
    \includegraphics[scale=1]{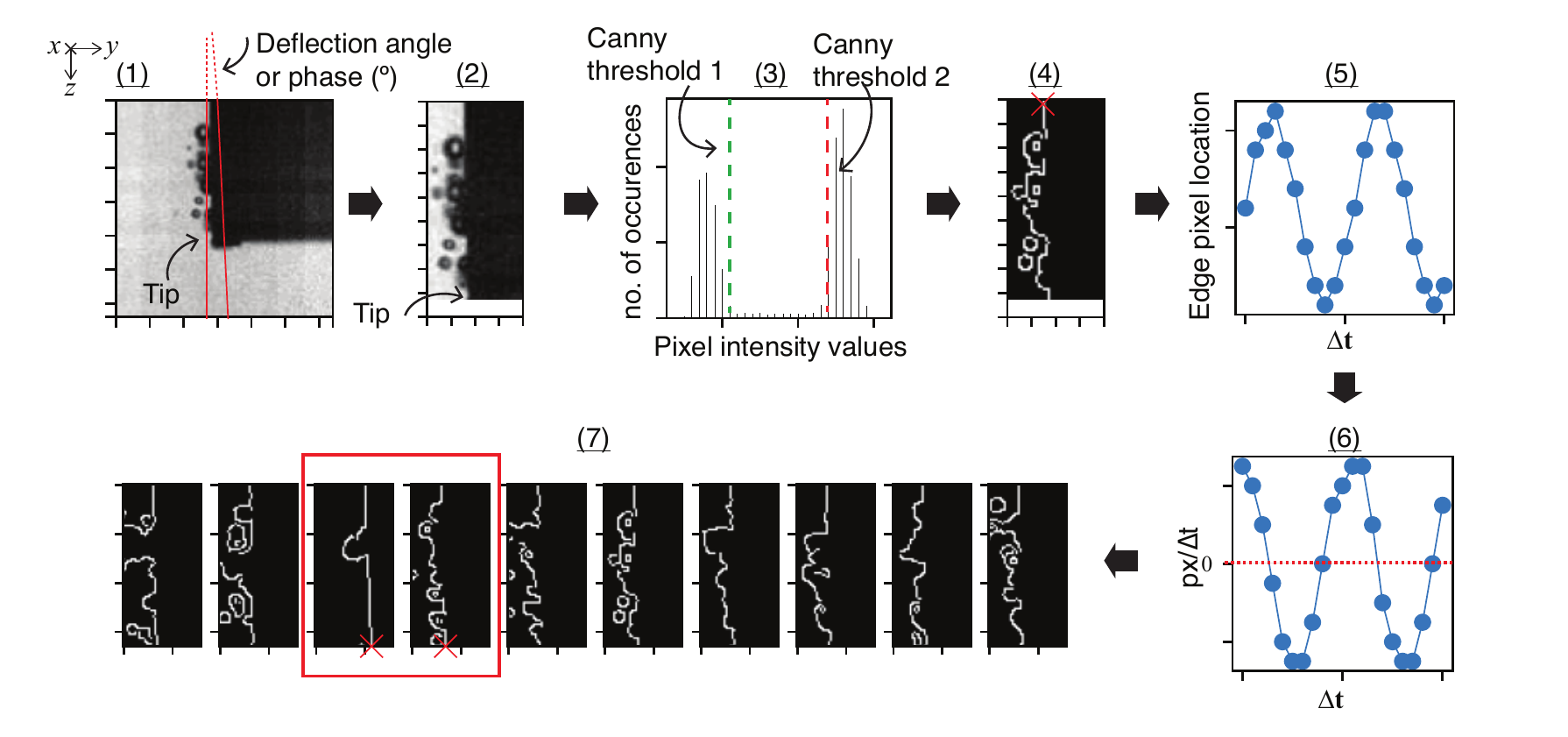}
    \caption{Needle tip deflection was measured using a sequence of frames captured from high-speed camera at 310 kHz, using a 7-step algorithm (1--7), involving cropping (1--2), Canny edge detection (3--4), computation of edge pixel location (5) and the derivative (6) in time, and finally measuring the peak-to-peak deflection at tip from a visually inspected frame pair (7).}
\label{fig:defsetup}
\end{figure*}

For each needle bevel type, we recorded 300 high speed camera frames, measuring 128 $\times$ 128 pixels with a spatial resolution of 1/180 mm ($\approx$ 5 \textmu m) per pixel, and a time resolution of 310 000 frames per second. As outlined in Fig.~\ref{fig:defsetup}, each frame (1) was cropped (2) so that needle tip was located in the last row (bottom) of frame, then the histogram of the image was computed (3), so that Canny thresholds 1 and 2 could be determined. Then Canny edge detection \cite{Canny1986} with a 3 $\times$ 3 Sobel operator was applied (4), and the location was computed for a cavitation-free bevel-edge pixel (marked $\mathbf{\times}$) for all 300 time steps. To determine the peak-to-peak deflection at the tip, the derivative (using a central difference algorithm) was calculated (6), and the frames containing the local extrema (i.e. peaks) of deflection were identified (7). Following a visual inspection for cavitation-free edges, a frame pair (or two frames that are half of the time period apart) was chosen (7), and the deflection at the tip was measured (marked $\mathbf{\times}$). The above was implemented in Python (v3.8, Python Software Foundation, python.org), utilising OpenCV's Canny edge detection algorithm (v4.5.1, Open Source Computer Vision Library, opencv.org).

\subsection*{Insertion media}
\label{Needle insertion-media}

Measurements were done in air (22.4--22.9$^{\circ}$ C), deionised water (20.8--21.5$^{\circ}$ C), and aqueous ballistic gelatin 10\% (w/v) (19.7--23.0$^{\circ}$ C, Honeywell\textsuperscript{\texttrademark} Fluka\textsuperscript{\texttrademark} Gelatin from bovine and porcine bones, for ballistic analysis type I, Honeywell International Inc., North Carolina, USA). Temperature was measured using a thermocouple type-K amplifier (AD595, Analog Devices Inc., Massachusetts, USA), coupled with a type-K thermocouple (Fluke 80PK-1 Bead Probe no. 3648 type-K, Fluke Corporation, Washington, USA). Depth was measured from the surface of medium (set as the origin of $z$-axis), using a vertical $z$-axis motorised translation stage (8MT50-100BS1-XYZ, Standa Ltd., Vilnius, Lithuania) with resolution of 5 \textmu m per step.

\subsection*{Statistical analysis}
Since the sample size was small (i.e. $n$ = 5), and normality could not be assumed, a two-sample, two-sided, Wilcoxon rank sum test was used, to compare the tip-deflection magnitudes of the different needle bevels. Multiple tests (3 tests per group) were conducted for each bevel group, so a \textit{Bonferroni}-correction was applied. The resulting $p$-value was 0.017, at 5\% error rate. The tests were conducted using R (v4.0.3, R Foundation for Statistical Computing, r-project.org).

\subsection*{Code availability}
The codes used for this study are available on reasonable request.

\section*{Results}

\subsection*{Flexural wavelength and transfer mechanical mobility}

\begin{figure*}
    \centering
    \includegraphics[scale=1]{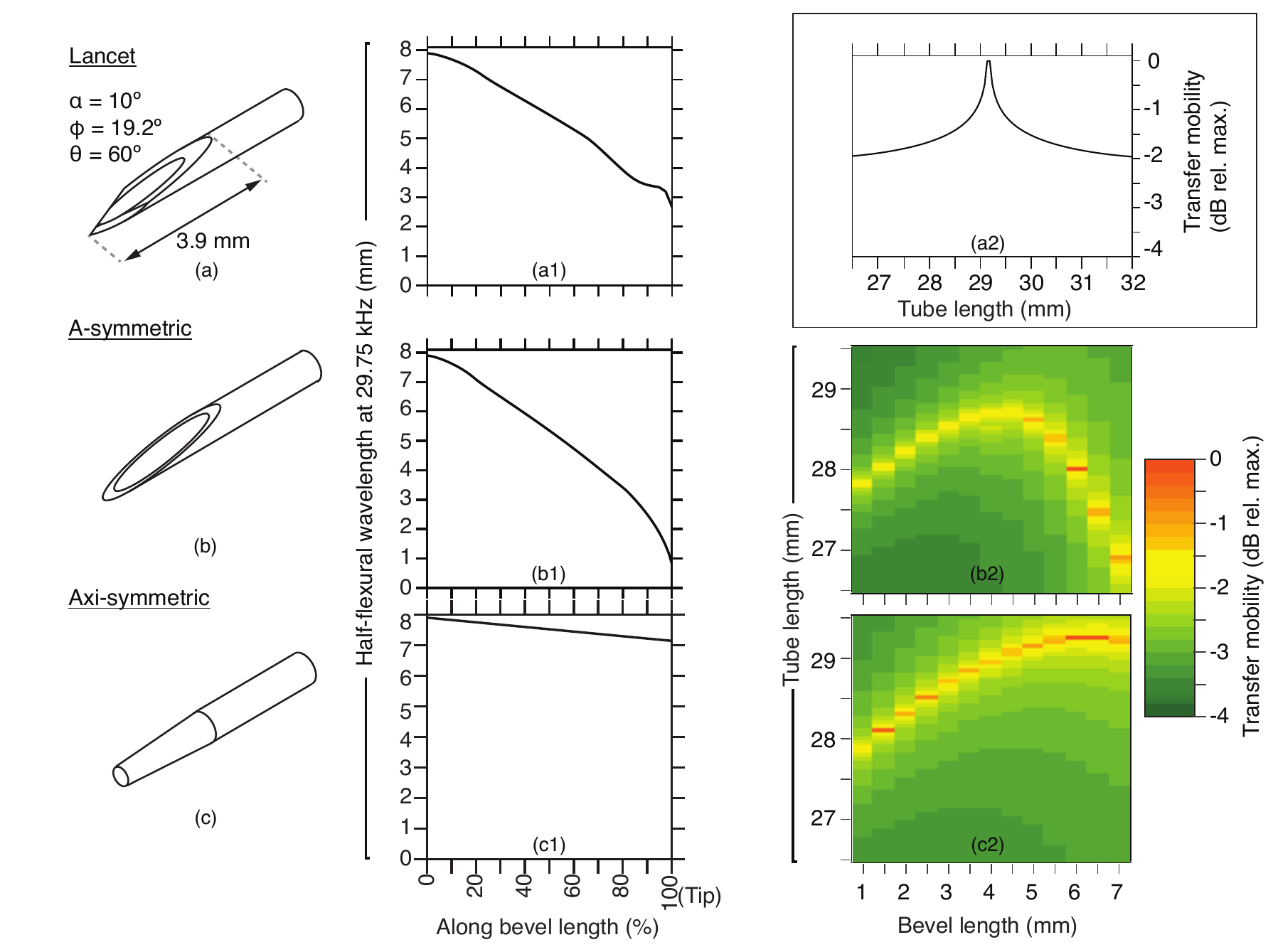}
    \caption{FEM studies at 29.75 kHz, revealed that the transfer mobility for axi-symmetric tip was less influenced by the change in tube length than its a-symmetric counterpart. \textbf{(a)} Lancet with specifications according to EN ISO 7864:2016, where $\alpha$ was the primary bevel angle, $\theta$ was the secondary bevel rotation angle, and $\phi$ was the secondary bevel angle when rotated, measured in degrees ($^\circ$). \textbf{(b)} Linear a-symmetric single step bevel (referred to as ‘standard’ in DIN 13097), and \textbf{(c)} linear axi-symmetric (circumferential) single-step bevel, with tip thickness of $\approx$ 50 \textmu m. \textbf{(a1, b1, c1)} Application of equation \eqref{flexwave} in calculation of the variation of flexural half-wavelength ($\lambda_y/2$) for bevel geometries (a), (b), and (c), using 21 Gauge needle tubing (0.80 mm outer diameter, 0.49 mm inner diameter, tube wall thickness 0.155 mm, regular wall, as specified in EN ISO 9626:2016), at frequency of 29.75 kHz, made of stainless steel grade 316 (\textit{Young}'s modulus 205 GN/m\textsuperscript{2}, density 8070 kg/m$^\textnormal{3}$, and \textit{Poisson}'s ratio 0.275). Mean $\lambda_y/2$ was 5.65, 5.17, and 7.52 mm for bevels (a), (b), and (c), respectively. \textbf{(a2, b2, c2)} Bevel length (BL) \textit{versus} tube length (TL) in a frequency domain study employing FEM (boundary conditions as in Fig.~\ref{fig:modelsetup}). The needle tube was excited flexurally at 29.75 kHz, and vibration was measured at the tip and presented as the magnitude of the transfer mechanical mobility (dB relative to maximum), for TLs 26.5--29.5 mm (step size 0.1 mm) and BLs 1--7 mm (step size 0.5 mm).}
\label{fig:FEMresults}
\end{figure*}

The following refers to Fig.~\ref{fig:FEMresults}. At 29.75 kHz, the flexural half-wavelength ($\lambda_y/2$) for 21 Gauge needle tubing was $\approx$ 8 mm. The flexural wavelength decreased along bevel when approaching tip. At tip, $\lambda_y/2$ was $\approx$ 3, 1, and 7 mm for the conventional lancet (a), a-symmetric (b), and axi-symmetric (c) single step bevels, respectively (Fig.~\ref{fig:FEMresults}(a, b, c)). Consequently this meant the range of the variation was $\approx$ 5 mm for the lancet (owing to the two lancet planes generating a single sharp point \cite{Wang2014}\textsuperscript{,} \cite{han2012models}), 7 mm for the a-symmetric bevel, and 1 mm for the axi-symmetric bevel (where the centre of gravity stayed constant, so effectively only tube wall thickness varied along bevel). Peaks of $\lvert\Tilde{Y}_{v_yF_y}\rvert$ indicated optimal TL-BL combinations (Fig.~\ref{fig:FEMresults}(a2, b2, c2)). For the lancet (a), since its dimensions were fixed, optimal TL was $\approx$ 29.1 mm. For the a-symmetric (b) and axi-symmetric (c) bevels, FEM studies included BLs 1 to 7 mm, so the optimal TLs varied from 26.9 to 28.7 mm (range 1.8) and 27.9 to 29.2 mm (range 1.3), respectively. For the a-symmetric bevel (b), optimal TLs increased steadily reaching a plateau at 4 mm BL, then steeply declined from 5 to 7 mm BL. For the axi-symmetric bevel (c), optimal TLs increased steadily with increasing BLs, and eventually plateaued at $\approx$ 6 to 7 mm BL.

\subsection*{Modal behaviour}

The needle construct exhibited three natural frequencies $f_{1-3}$, which were categorised into low, middle and high modal regions, as summarised in Table.~\ref{tab:ModalRegions}. The following refers to (Fig.~\ref{fig:DSPAnalysisResults}). 

\textbf{1$^\textnormal{st}$ Modal Region}: $f_1$ did not vary greatly with type of insertion medium, but varied with changing bevel geometry. $f_1$ decreased with decreasing bevel length (27.1, 26.2, and 25.9 kHz for AX1--3, in air, respectively). The region-averages of PTE\textsubscript{1} and FWHM\textsubscript{1} were $\approx$ 81\% and 230 Hz, respectively. FWHM\textsubscript{1} was highest in gelatin for the Lancet (L, 473 Hz). Note it was not possible to estimate FWHM\textsubscript{1} for AX2 in gelatin, due to low recorded magnitudes of FRF.

\textbf{2$^\textnormal{nd}$ Modal Region}: $f_2$ varied with type of insertion medium and bevel. In air, water, and gelatin, averages of $f_2$ were 29.1, 27.9, and 28.5 kHz, respectively. This modal region also exhibited PTE as high as 99\%, which was the highest among all measurement groups, with a region-average of 84\%. The region-average of FWHM\textsubscript{2} was $\approx$ 910 Hz.

\textbf{3\textsuperscript{rd} Modal Region}: $f_3$ frequencies varied with type of insertion medium and bevel. In air, water, and gelatin, the average values of $f_3$ were 32.0, 31.0, and 31.3 kHz, respectively. The region-average of PTE\textsubscript{3} was $\approx$ 74\%, which was the lowest among all regions. The region-average of FWHM\textsubscript{3} was $\approx$ 1085 Hz, which was higher than 1\textsuperscript{st} and 2\textsuperscript{nd} regions.

\subsection*{Measured deflection}

\begin{figure}
    \centerline{\includegraphics[scale=1]{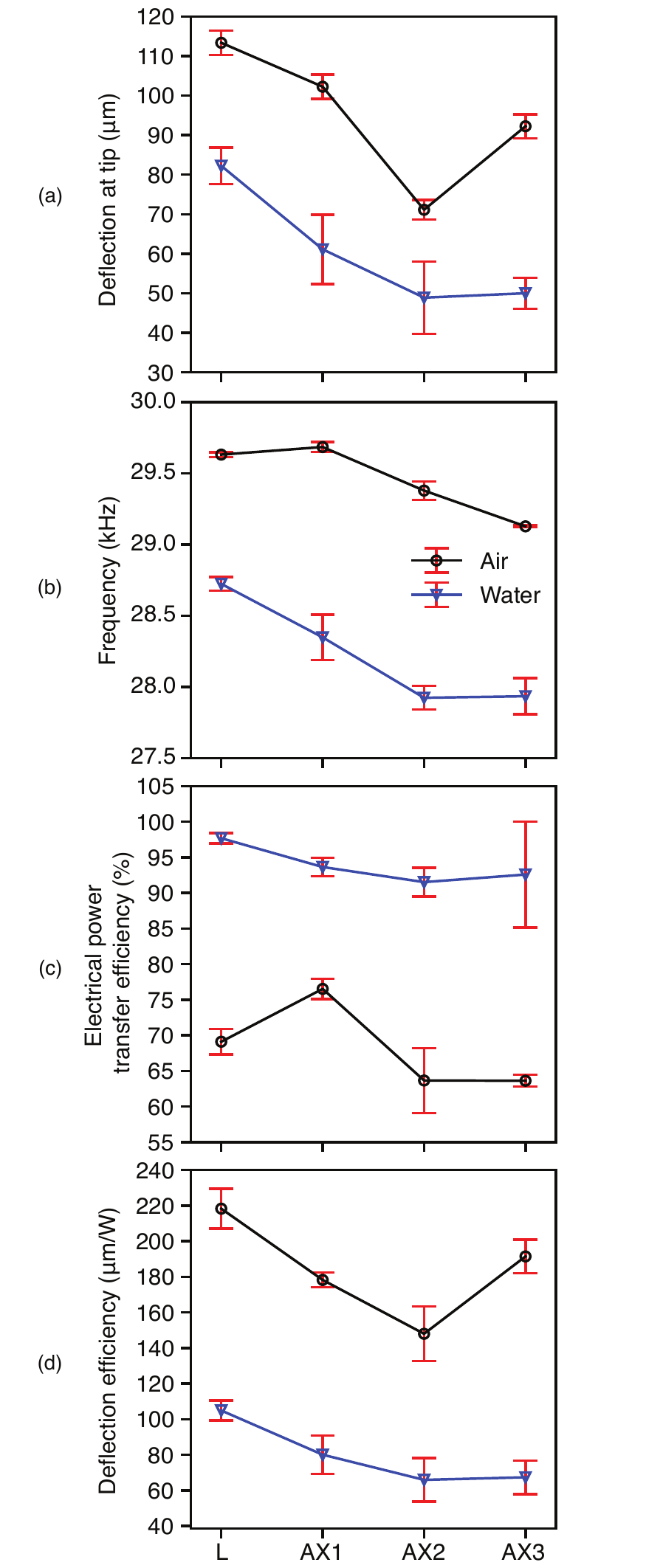}}
    \caption{Measured flexural-deflection magnitudes of needle tip (mean $\pm$ s.d., $n$ = 5) of bevels L and AX1--3 in air and water (20 mm depth), revealed the effects of changing bevel geometry. Measurements were obtained using continuous single-frequency sinusoid excitation. (\textbf{a}) Peak-to-peak deflection ($u_y\Vec{j}$) at the tip-point, measured at (\textbf{b}) their respective modal frequencies $f_2$. (\textbf{c}) The electrical power transfer efficiency (PTE, r.m.s., \%) as in equation \eqref{alpha_t}, and (\textbf{d}) deflection efficiency (\textmu m/W) was calculated as the ratio of the peak-to-peak deflection over the transmitted electrical power $P_T$.}
\label{fig:def_results}
\end{figure}

\begin{figure} 
    \centering
    \includegraphics[scale=1]{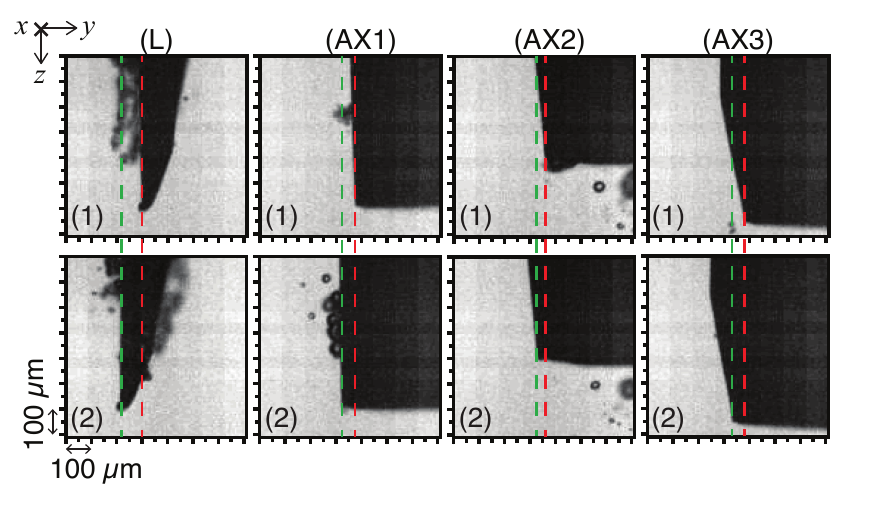}
    \caption{Typical high-speed camera shadowgraphs showing the peak-to-peak tip deflection (green and red dashed lines) for the lancet (L) and axi-symmetric tips (AX1--3), in water (depth 20 mm), during a half-cycle, at excitation frequency $f_2$ (sampling frequency 310 kHz). Captured greyscale images measured 128 $\times$ 128 pixels and the spatial resolution was $\approx$ 5 \textmu m per pixel.}
    \label{fig:WaterDef}
\end{figure}

The following refers to Fig.~\ref{fig:def_results} and Table.~\ref{tab:ST}. The lancet (L) deflected the most (with high significance to all tips, $p <$ 0.017) in both air and water (Fig.~\ref{fig:def_results}(a)), achieving the highest deflection efficiency (up to 220 \textmu m/W in air). In air, AX1 which had had higher BL, deflected higher than AX2--3 (with significance, $p <$ 0.017), while AX3 (which had lowest BL) deflected more than AX2 with deflection efficiency of 190 \textmu m/W. In water at 20 mm, no significant differences ($p >$ 0.017) were found in deflection and PTE for AX1--3. PTE levels were overall higher in water (90.2--98.4\%) than air (56--77.5\%) (Fig.~\ref{fig:def_results}(c)), noting cavitation events were clearly present in water during experimentation (Fig.~\ref{fig:WaterDef}).

\section*{Discussion}
\label{sec:discussion}

\begin{figure*}
    \centering
    \includegraphics[scale=1]{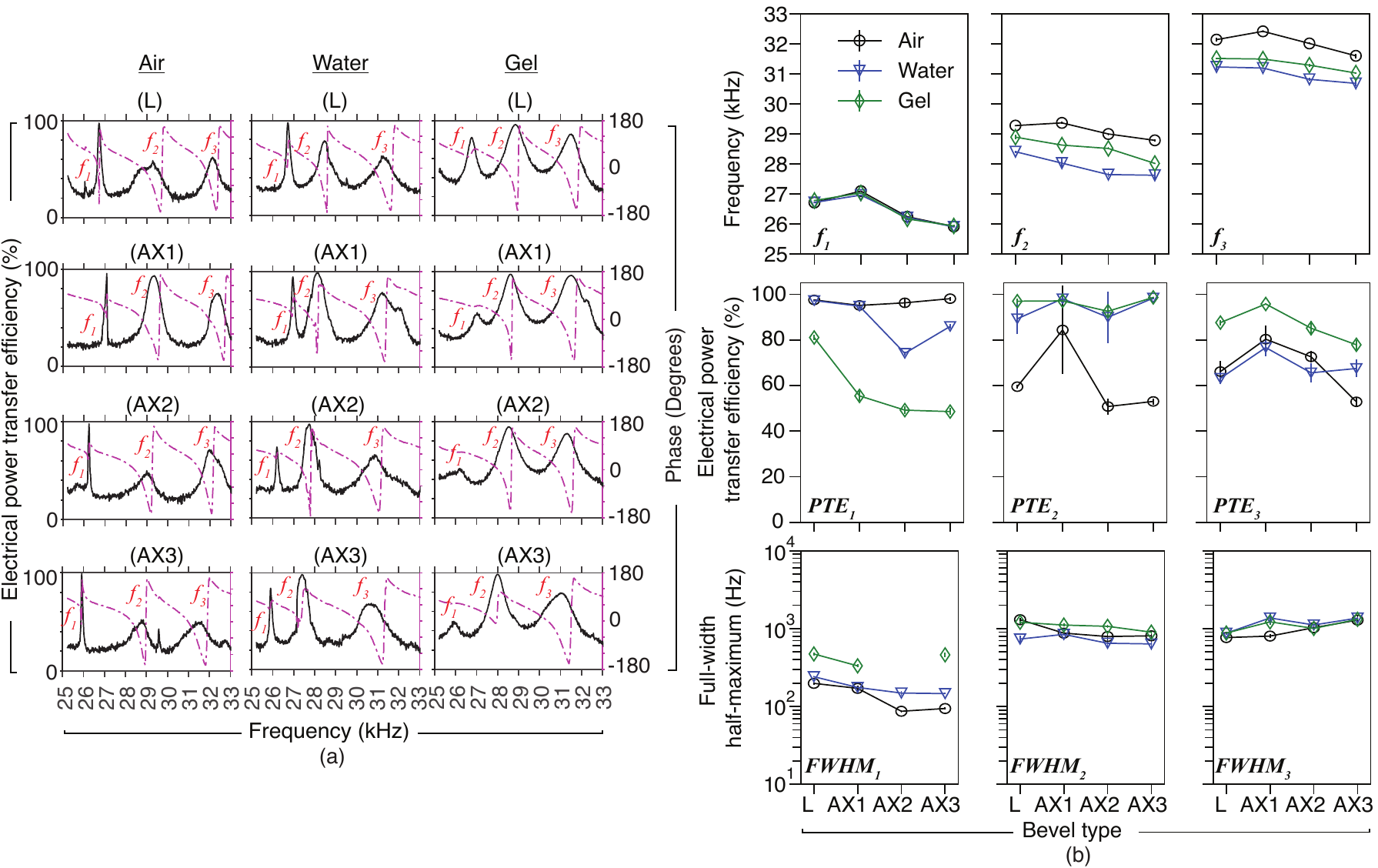}
    \caption{Mode $f_2$ was considered to be the most appropriate for comparing deflection of bevels, since it exhibited the highest levels of power transfer efficiency ($PTE_2$), which were as high as 99\%. \textbf{(a)} Typical recorded magnitudes of the instantaneous PTE obtained using swept-sine excitation, for the lancet (L) and axi-symmetric bevels AX1--3, in air, water, and and gelatin, at depth of 20 mm. One-sided spectra are shown. Measured FRFs (sampling frequency 300 kHz) were low-pass filtered, then downsampled by a factor of 200, for the purpose of modal analysis. The signal-to-noise ratio was $\leq$ 45 dB. Phase (dashed purple line) of PTE is shown in degrees ($^{\circ}$).  \textbf{(b)} Analysis of the modal response (mean $\pm$ 1 s.d., $n$ = 5), for bevels L and AX1--3) in air, water and gelatin 10\% (depth 20 mm), featuring (\textbf{top}) three modal regions (low, middle and high), and their corresponding modal frequencies $f_{1-3}$ (kHz), (\textbf{middle}) power efficiency $PTE_{1-3}$ calculated using equation \eqref{alpha_t}, and (\textbf{bottom}) the full-width at half-maximum measurements FWHM\textsubscript{1--3} (Hz), respectively. Note measurement of bandwidth was omitted when low PTE were recorded, i.e. FWHM\textsubscript{1}, AX2. }
    \label{fig:DSPAnalysisResults}
\end{figure*}

In this study, we modelled the change of flexural wavelength and the effect of tube length for a conventional lancet, a-symmetric and axi-symmetric bevel geometries (Fig.~\ref{fig:FEMresults}). We then characterised the frequency behaviour for the conventional lancet and the three fabricated axi-symmetric bevels, in air, water and ballistic gelatin 10\% (w/v) (Fig.~\ref{fig:DSPAnalysisResults}), and identified the mode most appropriate for comparing deflection of bevels. Finally, we measured the flexural-wave deflection at the needle-tip in air and at 20 mm depth in water, and quantified the electrical power transfer efficiency to insertion medium (PTE, \%) and the deflection efficiency (\textmu m/W) for each bevel type (Fig.~\ref{fig:def_results}).

The simulation study with varying tube lengths, revealed that the transfer mechanical mobility of the axi-symmetric bevel, was less influenced by the change in tube length than its a-symmetric counterpart. For a-symmetric and axi-symmetric bevels, the length-dependency of mobility plateaued approx. at 4 to 4.5 mm and at 6 to 7 mm, respectively (Fig.~\ref{fig:FEMresults}(b2, c2)). This indicated that the bevel had a range of lengths providing similar mobility. The practical relevance of this finding translates to manufacturing tolerance of the bevel length or flexibility to choose an appropriate frequency to achieve similar mobility.

In the experimental studies, the magnitude of the reflected flexural waves were affected by the boundary conditions at the needle-tip. When the needle-tip was inserted in water and gelatin, the average of PTE\textsubscript{2} was $\approx$ 95\%, compared to an average of 73\% and 77\% for PTE\textsubscript{1} and PTE\textsubscript{3}, respectively (Fig.~\ref{fig:DSPAnalysisResults}(b)). This suggested that the greatest transmission of acoustic energy into the embedding medium, i.e. water or gelatin, occurred at $f_2$. Similar behaviour was observed in a previous study \cite{Ying2006} with a simpler device construct at 41--43 kHz, where the authors showed the voltage reflection coefficient related to the mechanical modulus of insertion medium. The additional value for this approach could be an avenue to develop ways to estimate mechanical modulus of the insertion medium, while operating USeFNAB device.

The results show that the needle-bevel geometry affects the deflection amplitude at the needle-tip. The lancet achieved the highest deflection, as well as the highest deflection efficiency, in comparison to axi-symmetric bevels, which on average deflected less  (Fig.~\ref{fig:def_results}). Measurements in water suggested that, in terms of the peak deflection at the tip, there was no clear benefit of having longer bevel lengths. This is because axi-symmetric 4 mm bevel (AX1) having the longest bevel length, achieved the highest deflection among other axi-symmetric needles (AX2--3), but the difference was only statistically significant ($p < 0.017$) in air (Table.~\ref{tab:ST}). Considering this, the results suggest that the bevel geometries investigated in this study have a larger effect on deflection amplitudes than the bevel length. This may be associated with the flexural moduli depending e.g. on the overall thickness of the flexurally-bending material.

The study included the following experimental limitations. The total length of the needle tube (i.e. from tip to hub, see Fig.~\ref{fig:bevels}(b)) was $\approx$ 0.4 cm longer for the lancet (L), than the other needles (AX1--3). This could have influenced the modal response of needle construct. In addition, the shape and volume of the soldering at the waveguide-needle termination (see Fig.~\ref{fig:bevels}), may have affected the mechanical impedance of the needle construct, introducing uncertainty in mechanical impedance and bending behavior.

To conclude, we have demonstrated experimentally that bevel-geometry affects deflection amplitudes in USeFNAB. We recorded deflections in air of up to 113 \textmu m for a conventional needle, and 102 \textmu m for an axi-symmetric needle having a 4 mm long bevel. A bevel geometry exhibiting higher deflections might be beneficial in influencing cutting mechanisms, which could contribute to influence the quantity and quality of biopsy samples. Ultimately, when designing an optimal bevel geometry and the mechanical modulus of target tissue type are of paramount importance to ensure optimal operation.

\section*{Data availability}

The datasets produced during this study are available from the corresponding author on reasonable request.

\bibliography{main}

\section*{Acknowledgements}

Business Finland (grant 5607/31/2018) and the Academy of Finland (grants 311586, 314286, and 335799) are acknowledged for financial support. The authors would like to give thanks to Yohann Le Bourlout and Dr. Gösta Ehnholm for fabrication and design of needle constructs, RF amplifier, and SWR meter. We are grateful to Jouni Rantanen, Dr. Maxime Fauconnier, Emanuele Perra, Dr. Balasubramanian Nallannan, and all other members of the Medical Ultrasonics Laboratory (MEDUSA) at Aalto University (Finland), for the insightful discussions. In addition, the authors would like to thank senior lecturer Kari Santaoja (Aalto University) for the constructive discussions.

\section*{Author contributions statement}

All authors have contributed to design of study, writing, and review of the manuscript. S.B. produced and analysed all of the data.

\section*{Competing interests}

S.B. has no competing interests. H.J.N. has stock ownership in Swan Cytologics Inc., Toronto, ON, Canada, and is an inventor within patent applications WO2018000102A1 and WO2020240084A1.

\clearpage
\newpage

\begin{table}
\centering
\caption{Three Modal Regions of needle constructs L, and AX1--3}
\setlength{\tabcolsep}{6pt}
\begin{tabular}{|c|c|p{108pt}|}
\hline
\textbf{Modal Region} & 
\textbf{Frequency (kHz)} & 
\textbf{Description} \\
\hline
1 & $25 \leq f_1 \leq 27.5$ & Low modal region \\
\hline
2 & $27.5 < f_2 \leq 30$ & Middle modal region \\
\hline
3 & $30 < f_3 \leq 33$ & High modal region \\
\hline
\end{tabular}
\label{tab:ModalRegions}
\end{table}

\begin{table}
\centering
\caption{Was there significant difference in tip-deflection between the lancet (L) and the axi-symmetric bevels (AX1--3)?\textsuperscript{a}}
\centering
\setlength{\tabcolsep}{3pt}
    \begin{tabular}[c]{|l|c|c|c|c|c|c|}
    \hline
    Test-Pair&L-AX1&L-AX2&L-AX3& AX1-AX2& AX1-AX3&AX2-AX3\\
    \hline
    \multicolumn{7}{p{246pt}}{Tests for deflection at tip ($u_y\Vec{j}$, Fig.~\ref{fig:def_results}(a))} \\
    \hline
    \textbf{Air} & 0.010 & 0.009 & 0.010 & 0.009 & 0.010 & 0.009 \\
    \hline 
    \textbf{Water} & 0.012 & 0.011 & 0.011 &\color{red} 0.087 &\color{red} 0.052 &\color{red} 0.821 \\
    \hline
    \multicolumn{7}{p{246pt}}{Tests for electrical power transfer efficiency (PTE, Fig.~\ref{fig:def_results}(c))} \\
    \hline
    \textbf{Air} & 0.008 & 0.016 & 0.008 & 0.008 & 0.008 &\color{red} 0.310 \\
    \hline
    \textbf{Water} & 0.010 & 0.010 &\color{red} 0.310 &\color{red} 0.095 &\color{red} 0.691 &\color{red} 0.691 \\
    \hline
    \multicolumn{7}{p{246pt}}{\textsuperscript{a} Each pair of bevels (e.g. L \textit{vs.} AX1) were tested using a two-sample, two-sided, Wilcoxon rank sum test, $n$ = 5, 5\% error rate, $p$-value was 0.017 (\textit{Bonferroni}-correction was applied). Red indicates no statistical significance found between bevel pair (i.e. $p$ $\geq$ 0.017).} \\
    \end{tabular}
\label{tab:ST}
\end{table}

\end{document}